\documentclass{article}
\usepackage{amssymb}

\def\thefootnote{\fnsymbol{footnote}}
\begin{document}
\begin{titlepage}
\today          \hfill 
\begin{center}
\hfill    LBNL-xxxxx \\
\hfill    UCB-PTH-xx/xx \\
\hfill hep-th/xxxxxxx \\

\vskip .5in
\renewcommand{\thefootnote}{\fnsymbol{footnote}}
{\Large \bf Unstable solitons on noncommutative tori and D-branes} 
\footnote{This work was supported by the Director, Office of Energy 
Research, Office of High Energy and Nuclear Physics, Division of High 
Energy Physics of the U.S. Department of Energy under Contract 
DE-AC03-76SF00098 and in part by the National Science Foundation grant PHY-0098840.}
\vskip .50in

\vskip .5in
{\large Anatoly Konechny}\footnote{email address: konechny@thsrv.lbl.gov}

\vskip 0.5cm
{\em Department of Physics\\
University of California at Berkeley\\
   and\\
 Theoretical Physics Group, Mail Stop 50A-5101\\
    Lawrence Berkeley National Laboratory\\
      Berkeley, California 94720}
\end{center}

\vskip .5in

\begin{abstract} \large
We describe a class of exact solutions of super Yang-Mills  theory on even-dimensional 
noncommutative tori.  These solutions  generalize the solitons on a noncommutative plane 
 introduced in \cite{unstable}
that are conjectured to describe unstable D2p-D0 systems. We show that the spectrum of 
quadratic fluctuations around our solutions correctly reproduces the string spectrum of  
the D2p-D0 system in the Seiberg-Witten decoupling limit.
In particular the fluctuations correctly  reproduce the 0-0 string winding modes.     
For $p=1$ and $p=2$  we match the differences  between the soliton energy and the energy of an
 appropriate SYM BPS state with the binding energies  of D2-D0 and D4-D0 systems. 
We also give an example of  a soliton that we conjecture describes branes of intermediate dimension 
on a torus such as a D2-D4 system on a four-torus.  
 
\end{abstract}
\end{titlepage}
\large

\newpage
\renewcommand{\thepage}{\arabic{page}}
\setcounter{page}{1}
\section{Introduction} 
The D0-D2p system in the presence of a constant background $B$-field was first studied in \cite{GNS}. 
Later in \cite{SW} it was shown that in a certain decoupling limit the low energy effective 
theory on D2p-branes with a constant $B$-field can be described by means of a noncommutative super Yang-Mills (SYM) 
theory. 
The D0-D2p system in a $B$-field  was  further analyzed in \cite{SW}, \cite{Murakami_etal}. 
For a generic constant $B$-field such systems with point-like D0-branes are unstable. 
However for a certain locus of the $B$-field moduli we can have bound states with smeared D0-branes and 
in  more restricted regions we can have bound states of point-like D0 branes sitting on top of 
the D2p-brane. Thus it was shown in \cite{Witten68} (see  also \cite{Park_etal}, cite{Sato}) 
that the D0-D6 system, which is not supersymmetric in the absence of a $B$-field background, becomes supersymmetric 
for certain values of $B$.     
 
In papers \cite{Polikr}, \cite{GrNek1}, \cite{Bak}, \cite{unstable}, \cite{exact}, \cite{GrossNek}  
a number of exact solitonic solutions in noncommutative SYM theory on a plane was  constructed and 
their interpretation in terms of D-branes was given. In particular in \cite{unstable} solitonic 
solutions describing unstable D0-D2 and D0-D4 systems were studied. It was shown that the solitons 
have a correct fluctuation spectrum and their energies match with the binding energies in the corresponding 
brane systems. These  exact solutions describing  non-BPS configurations are  powerful tools 
in studying such systems, in particular tachyon condensation.

The aim of the present paper is to construct analogous solitonic solutions on tori in the presence 
of a large $B$-field background (understood in terms of the decoupling limit of \cite{SW}).  
To describe our approach to constructing such solitons let us first remind the reader the essential 
ingredients in the construction of SYM solitons on the plane as presented in the papers cited above.

First, our fields are functions on a noncommutative plane and they can be conveniently represented as 
operators acting on a Fock space.  More precisely in  complex coordinates on a two-plane (taken just for the purposes of 
illustration)  one has 
the following representation of gauge fields (here we use the notations of \cite{unstable})
$$
A_{z} = iC - a^{\dagger} \, , \qquad A_{\bar z}= -i\bar C - a
$$ 
where $a$ and $a^{\dagger}$ are annihilation and creation operators and $C$ and $\bar C$ are essentially 
arbitrary operators 
(to specify a finite energy  gauge field the fields $A_{z}$, $A_{\bar z}$  have to be gauge trivial at infinity, see 
\cite{Schwarz} for a detailed discussion). 
In terms of the variables $C$, $\bar C$ the Yang-Mills equations of motion in the Coulomb gauge  take the form
$$
\partial_{t}^{2}C=[C,[C,\bar C]]
$$
plus the conjugated equation. 

Next one can introduce the so called partial isometry operators 
$S$ and $S^{\dagger}$ that satisfy 
$$
SS^{\dagger}=1 \, , \quad S^{\dagger}S=1 - P_{0}
$$   
where $P_{0}$ is a projector onto the Fock vacuum subspace.
It is easy to see then that dressing transformations 
$$
C \mapsto (S^{\dagger})^{m}a^{\dagger}S^{m} 
$$
generate solutions of the static equation $[C,[C,\bar C]]=0$ and carry  $m>0$ units of magnetic flux.

Such a dressing transformation singles out a finite dimensional subspace in our Fock space. 
The projector on this subspace is $P_{m-1} = \sum_{i=0}^{m-1}|i\rangle\langle i|$. 
The curvature operator has nonvanishing components only for matrix elements between vectors from this 
subspace. If we consider fluctuations around our soliton then fluctuations with matrix elements between vectors from the 
subspace singled out by  $P_{m-1}$   describe strings attached to a D0 brane (that in this case seats at the origin). Fluctuations  
 with matrix elements from the orthogonal complement  describe D2-D2 strings and 
the off-diagonal ones describe D0-D2 strings.

As it can be easily deduced from this construction (see  \cite{review2}  sections 3.1, 3.2 for a pedagogical 
explanation)  a generic  module (representation)
  of the algebra of functions  on a noncommutative plane denoted ${\mathbb R}_{\theta}$ has a form of a direct sum
$$
({\cal F})^{m}\oplus ({\mathbb R}_{\theta})^{n} 
$$
where $\cal F$ is a module that carries a unit of magnetic flux and is built on a Fock representation 
of the operator algebra  ${\mathbb R}_{\theta}$ and  $ ({\mathbb R}_{\theta})^{n} $ stands for 
$n$-copies of a free module, i.e. a copy of the algebra itself  with an action  by a (right or left) 
multiplication. This can be shown to hold for an arbitrary even-dimensional noncommutative plane
 with the numbers $m$ and $n$ being analogs of the instanton number and the rank of the gauge group, 
or respectively D0 and D2 brane charges.

The projector $P_{m-1}$ can be viewed as a projector specifying a projective module, that is 
an analog of vector bundles in noncommutative geometry. For  noncommutative tori  
such projectors specifying projective modules that carry nontrivial topological numbers are also known. 
In the two-dimensional case there is an explicit construction by Powers and Rieffel 
\cite{Cst}. There are also other approaches to constructing projectors based on  
theta-functions \cite{Boca}, \cite{MM}. These constructions  are certainly  more complicated then the simple 
expression for the $P_{m-1}$ above  and we find them  to be no very convenient 
from the computational  point of view. On the other hand there is a different, more straightforward description 
of projective modules over noncommutative tori \cite{RieffelProj} when one simply specifies an action 
of the algebra  generators on a Hilbert space. 
There is a simple class of projective modules that posses a constant curvature connection. These modules 
and connections on them  can be used as 
building blocks to construct modules with various topological numbers and solitonic solutions on them. 
 This will be our starting point  in constructing solitons describing unstable D-brane systems on tori. 
Despite the fact that we will not be using explicit projector operators  and partial isometries we 
will still be able effectively describe those solitons, compute their energy and fluctuations and 
analyze their moduli. We will see however (see the last section of the paper) that this approach has 
some limitations and contrary to the unstable ones the BPS solutions can have a complicated description 
in the picture we use. And it would be desirable for the purposes of describing tachyon condensation to 
construct some analogs of partial isometry operators.

The paper is organized as follows. Our main example will be D0-D2 system on a two-torus. 
In section 2 we will describe projective modules describing a single D2, a single D0 and a D0-D2 system. 
In section 3 we introduce a solitonic solution on the D0-D2 module. We match the excess energy of the 
soliton above the energy of the SYM BPS state with the binding energy of the D0-D2 system, 
match the spectrum of fluctuations and discuss the solitons moduli space.
In section 4 we introduce various generalizations of the system discussed in section 2. 
First we introduce a  soliton on an arbitrary even-dimensional torus that generalizes the D0-D2 soliton. 
We compute the difference between the soliton energy and the SYM energy of the BPS state for the case of a four-torus 
and match it with the D0-D4 binding energy. We also show how one can construct  solitons  that we conjecture to 
describe a system with branes of intermediate dimension. We pick a D2-D4 system on a four-torus to illustrate this 
type of generalization. Finally, in that section  we also discuss systems with multiples of branes of various dimension. 
In  section 5 we discuss some open problems, in particular application of our results to tachyon condensation.


\section{D2, D0 and D0-D2 modules} 
In this section we will introduce the  basic mathematical constructions needed to describe D0 and D2 branes 
on noncommutative two-tori. We will be rather succinct and will only introduce the objects directly needed to describe 
the D0-D2 system. We refer the reader to the review papers \cite{review1}, \cite{review2} for a more 
detailed exposition of the mathematical tools used here.

Consider a torus $T^{d}$ with coordinates $x^{i}\sim x^{i} +1$, $i=1, \dots , d$ and metric $G_{ij}$. 
Given an antisymmetric $d\times d$ tensor $\theta^{ij}$ we can define an algebra of functions on a 
 noncommutative torus $T_{\theta}^{d}$ as an algebra with generators $U_{i}$ satisfying the relations 
\begin{equation}
U_{i}U_{j} = e^{2\pi i \theta^{ij}}U_{j}U_{i} \, . 
\end{equation}
It is convenient to define linear generators 
\begin{equation} \label{Un}
U_{\bf n} = U_{1}^{n_{1}} U_{2}^{n_{2}} \cdot \dots \cdot U_{d}^{n_{d}}e^{-\pi i \sum_{j<k}n_{j}n_{k}\theta^{jk}}
\end{equation}
where ${\bf n} = (n_{1}, \dots , n_{d})$, $n_{i}$ are all integers and $U_{i}^{-n}$, $n_{i}>0$ is understood as 
an $n$-th power of an inverse operator. A general function on  $T_{\theta}^{d}$ can be represented as 
$$
f = \sum_{{\bf n}\in {\mathbb Z}^{d}} c({\bf n}) U_{\bf n} 
$$   
where $c({\bf n})$ are ${\mathbb C}$-valued coefficients.

Vector bundles over noncommutative tori are described by means of projective modules. 
Throughout the paper we work with right modules. 
To make the discussion more explicit in the remaining part of   this section we will  concentrate on two-dimensional tori. 
All of the constructions we will present below have a straightforward generalization for higher-dimensional tori to be 
discussed in section \ref{h.d.}.

We can assume that a two-dimensional noncommutative torus is 
specified  by a real number $\theta \ge 0$ so that $\theta^{ij} = \theta \epsilon^{ij}$. It is known that a projective module 
$E$ over $T_{\theta}^{2}$ is specified  by two integers $n, m$ such that ${\rm dim}(E) = n + m\theta >0$ \cite{ConnesRieffel}, 
\cite{RieffelProj}. 
These numbers specify a K-theory class of $E$ so that $n$ corresponds to the D2-brane charge and $M$ to the D0-brane charge respectively. 
Below we describe two elementary modules, one carrying a unit of D2 brane charge and another one carrying a unit of D0 charge.   
Both modules were discussed informally in section 6.2 of \cite{SW}. 

\noindent \underline{\bf D2 module.}
This module to be denoted  $E_{1, 0}$ has topological numbers $n=1$, $m=0$ and is just a rank one free module. 
 Its  elements  are elements of 
the  $T_{\theta}^{2}$ itself and the algebra acts  by multiplication from the right.
An arbitrary  Yang-Mills field on  $E_{1, 0}$ can be written in the form $\partial_{j} + i u_{j}$ where 
$\partial_{j}$ act according to  $\partial_{j}(U_{\bf n}) = 2\pi i n_{j}U_{\bf n}$ and $u_{j}$ are operators of 
left multiplication by  elements  $u_{j}\in T_{\theta}^{2}$ denoted by the same letters. We also assume that $u_{j}$ 
are Hermitian with respect to the involution $\ast$ that acts on the generators as $U_{\bf n}^{*} = U_{-{\bf n}}$.

\noindent \underline{\bf D0 module.} 
This module denoted $E_{0,1}$ has  topological numbers  $n=0$ and $m=1$. It can be explicitly defined in terms of 
 generators $U_{1}$, $U_{2}$ acting on functions of one variable  $\phi(x)\in L_{2}({\mathbb R}^{1})$ as follows 
\begin{equation} \label{Ui}
\phi(x)U_{1} = \phi(x - \theta) \nonumber \, , \qquad
\phi(x)U_{2} = e^{2\pi ix}\phi(x) \, .     
\end{equation}

The module  $E_{0, 1}$ has a constant curvature Yang-Mills field denoted  $\tilde \nabla_{j}^{0}$ that acts as 
\begin{equation} \label{ccc}
\tilde \nabla_{1}^{0}\phi(x) = \frac{2\pi i }{\theta}x\cdot \phi(x)\, , \qquad 
\tilde \nabla_{2}^{0}\phi(x) = \frac{\partial \phi(x)}{\partial x} \, ,
\end{equation}
and the curvature is 
\begin{equation}
F_{12} = [\tilde \nabla_{1}^{0}, \tilde \nabla_{2}^{0}] = -\frac{2\pi i }{\theta}\cdot {\bf 1}
\end{equation}
where $\bf 1$ stands for a unit operator on $E_{0,1}$.

An arbitrary connection can be represented in the form $\nabla_{j} = \tilde \nabla_{j}^{0} + i z_{j}$ 
where $z_{j}$ belongs to the algebra $End_{T_{\theta}^{2}}(E_{0,1})$ of endomorphisms of the module $E_{0,1}$. It 
consists of all operators acting on   $E_{0,1}$ that commute with the action of $T_{\theta}^{2}$. 
The algebra  $End_{T_{\theta}^{2}}(E_{0,1})$ is itself an algebra of functions on the T-dual noncommutative torus 
$T_{1/\theta}$. Its generators $Z_{j}$ act on $E_{0,1}$ as follows 
\begin{equation} \label{Zi}
Z_{1}\phi(x) = \phi(x + 1) \, , \qquad Z_{2}\phi(x) = e^{\frac{2\pi ix}{\theta}}\phi(x) \, .
\end{equation} 
It is straightforward to check that the operators $Z_{i}$ commute with the operators $U_{i}$ given in (\ref{Ui}) 
and satisfy the commutation relation 
$$
Z_{1}Z_{2}= e^{\frac{2\pi i }{\theta}}Z_{2}Z_{1} \, . 
$$
Similarly to generators $U_{\bf n}$ (\ref{Un}) we can introduce a set of linear generators $Z_{\bf n}$ for 
 $T_{1/\theta}^{2}$; one just needs to replace $\theta$ with $1/\theta$ and $U$'s with $Z$'s in (\ref{Un}). 
 Then for an arbitrary connection $ \tilde \nabla_{j}^{0} + i z_{j}$ we can write  
$$
z_{j} =  \sum_{{\bf n}\in {\mathbb Z}^{2}} a_{j}({\bf n}) Z_{\bf n} 
$$
where $a_{j}({\bf n})$ are numbers. 
(As a point of mathematical  rigor we should  remark here 
that the functions $\phi(x)$ should be assumed to be smooth and decaying at infinity faster than any 
polynomial to ensure a well defined action of operators $\tilde \nabla_{j}^{0}$ and polynomials thereof.) 

\noindent \underline{ \bf Fock space picture.} 
Note that the operators (\ref{ccc}) give a single copy of the irreducible representation of the Heisenberg algebra. 
The whole module $E_{0,1}$ can be described in terms of a Fock space that is very similar in spirit to the construction 
of D0-module on noncommutative ${\mathbb R}^{2}$. Introduce creation and annihilation operators as 
\begin{eqnarray} \label{aa}
a^{\dagger} & =& \sqrt{\frac{\pi}{\theta}}x -\sqrt{\frac{\theta}{4\pi}}\frac{\partial}{\partial x} \nonumber \\
a &=& \sqrt{\frac{\pi}{\theta}}x +\sqrt{\frac{\theta}{4\pi}}\frac{\partial}{\partial x}
\end{eqnarray} 
then we have 
\begin{equation} \label{n-a}
\tilde \nabla_{1}^{0}=i\sqrt{\frac{\pi}{\theta}}(a + a^{\dagger}) \, , \qquad 
\tilde \nabla_{2}^{0}=\sqrt{\frac{\pi}{\theta}}(a - a^{\dagger}) \, . 
\end{equation}
The generators $U_{j}$ of the torus and the generators $Z_{j}$ of the dual torus can also be expressed 
in terms of $a$ and $a^{\dagger}$ and for the vectors of $E_{0,1}$ instead of the Schroedinger representation 
$\phi(x)$ we can go to the occupation number representation.

\noindent \underline{\bf D0-D2  module.}
In the next section we will work with a module $E=E_{0,1}\oplus E_{1,0}$ that is a direct sum of the two modules discussed above. 
Elements of this module can be represented as a pair $(\phi(x), a)$, $\phi(x) \in  L_{2}({\mathbb R}^{1})$, $a\in T_{\theta}^{2}$ 
that we will right as a column vector
$$
\left( \begin{array}{c} \phi(x) \\
a \end{array} \right) \, .  
$$

An arbitrary Yang-Mills field on such a module can be written in the following form 
\begin{equation} \label{c2}
\nabla_{j} = \nabla_{j}^{0}  + i\left( 
\begin{array}{cc}
z_{j} & \psi_{j} \\
\psi^{*}_{j} & u_{j} 
\end{array} \right) 
 \end{equation}
where 
$$
 \nabla_{j}^{0} = 
\left( 
\begin{array}{cc}
\tilde \nabla^{0}_{j} & 0 \\
0 & \partial_{j}  
\end{array} \right)  
$$
is simply the direct sum of the two fiducial connections introduced  above. 
As about the second term in (\ref{c2}) the diagonal blocks $z_{j}$ and $u_{j}$ were already described above,  
the off-diagonal blocks $\psi_{j}$ and $\psi^{*}_{j}$ should specify  $T_{\theta}$-linear mappings from 
the free module $E_{1,0}$ to $E_{0,1}$ and in the opposite direction respectively. It follows from  $T_{\theta}$-linearity 
that $\psi_{j}(a) = \psi_{j}(1\cdot a) =\psi_{j}(1)a$ and hence to specify the mapping $\psi_{j}$ we only need to know its value 
on the unit element $1\in T_{\theta}^{2}$. The last one is a function from  $L_{2}({\mathbb R}^{1})$ to be denoted $\psi_{j}(x)$. 
The construction of the conjugated mappings $\psi^{*}_{j}: E_{0,1} \to E_{1,0}  $ is a bit more involved. Given a function 
$\phi(x)$ we need to map it into an  element of the torus algebra in a way that commutes with the action of $T_{\theta}^{2}$. 
This is done by using the    $T_{\theta}^{2}$-valued  inner product $\langle , \rangle_{T_{\theta}^{2}}$: 
$$
\psi^{*}_{j}(\phi ) = \langle \psi_{j} , \phi\rangle_{T_{\theta}^{2}}   
$$ 
where on the right hand side $\psi_{j}=\psi_{j}(x)$, $\phi=\phi(x)$ are both elements of $E_{0,1}$.
The inner product at hand can be defined by the following explicit formula 
\begin{eqnarray} \label{inner_pr}
&& \langle \psi , \phi\rangle_{T_{\theta}^{2}}= \sum_{{\bf n}\in {\mathbb Z}^{2}} \langle \psi, \chi U_{- \bf n}\rangle_{\mathbb C} 
U_{\bf n} = \nonumber \\
&&  \sum_{n_{1}, n_{2}} U_{\bf n} e^{-\pi i n_{1}n_{2}\theta}\int dx\,  \bar \psi(x)\phi(x + n_{1}\theta)e^{-2\pi ixn_{2}} \, .    
\end{eqnarray}
It is easy to check that this inner product satisfies the desired $T_{\theta}^{2}$-linearity property
$$
\langle \psi , \phi a\rangle_{T_{\theta}^{2}} = \langle \psi , \phi \rangle_{T_{\theta}^{2}}\cdot a 
$$
where on the LHS $a$ stands for the right action of  $a\in T_{\theta}^{2}$ on $E_{0,1}$ specified in  (\ref{Ui}) and on the 
RHS we have a multiplication of two elements of $T_{\theta}^{2}$.

According to the discussion in section 6.3 of  \cite{SW} the algebra $T_{\theta}^{2}$ is the algebra of ground state 2-2 strings, 
$T_{1/\theta}^{2}$ is that of 0-0 strings and they both act naturally on the 0-2 strings, each at the 
corresponding end of the string. The 0-2 strings  in their turn are  described by elements of 
$E_{0,1} \cong L_{2}({\mathbb R}^{1})$ and the actions of 2-2 and 0-0 algebras are given by operators (\ref{Ui}), (\ref{Zi}). The fact that we 
have a left and  a right action stems from the difference in outward orientations of the open string at $\sigma =0$ and 
$\sigma=\pi$. On the Yang-Mills side all of this fits very well with the interpretation of the diagonal blocks in (\ref{c2})     
as the ones describing excitations of 0-0 and 2-2 strings and the off-diagonal ones describing excitations of 0-2 strings.

In order to construct a SYM action functional on $E_{0,1}\oplus E_{1,0}$ we need to define a trace on the algebra 
of endomorphisms of this module, i.e. on operators having a $2\times 2$ block form  as in (\ref{c2}). It is defined as 
\begin{equation}
{\rm Tr} \left( 
\begin{array}{cc}
z & \psi \\
\psi^{*} & u 
\end{array} \right) 
 = {\rm Tr}_{1/\theta} z + {\rm Tr}_{\theta} u 
\end{equation} 
where ${\rm Tr}_{1/\theta}$ and ${\rm Tr}_{\theta}$ are traces on the algebras $T_{1/\theta}^{2}$ and $T_{\theta}^{2}$ 
respectively defined as 
\begin{eqnarray}
&& {\rm Tr}_{1/\theta}  \sum_{{\bf n}\in {\mathbb Z}^{2}} a({\bf n}) Z_{\bf n} = \theta a(0) \nonumber \\
&& {\rm Tr}_{\theta}  \sum_{{\bf n}\in {\mathbb Z}^{2}} c({\bf n}) U_{\bf n} =  c(0)
\end{eqnarray}  
(see e.g. \cite{review1} for an explanation of   the normalizations of the traces).
When we multiply two operators having a block form as above and then take a trace  the following identities \cite{Cst}, 
\cite{RieffelProj}
are quite useful 
\begin{equation} \label{identity}
{\rm Tr}_{1/ \theta} \psi \circ \chi^{*} = {\rm Tr}_{\theta} \chi^{*}  \circ \psi = 
{\rm Tr}_{\theta} \langle \chi, \psi \rangle_{T_{\theta}^{2}} = \langle \chi, \psi \rangle_{\mathbb C} \, . 
\end{equation}


\section{Solitons on a two-torus and D0-D2 system} 
\noindent \underline{\bf The soliton.}
We consider a SYM theory on $T_{\theta}^{d}\times {\mathbb R}^{1}$ where ${\mathbb R}^{1}$ corresponds to 
the time direction. The action functional in the Coulomb gauge 
has the form 
\begin{eqnarray} \label{action} 
S& =&- \frac{\sqrt{G}}{4g_{YM}^{2}} {\rm Tr} \Bigl( -2\partial_{t}A_{i}G^{ij}\partial_{t}A_{j} +        F_{ij}F_{kl}G^{ik}G^{jl} +  
2[\nabla_{i}, X_{I}]G^{ij}[\nabla_{j}, X^{I}] + [X_{I}, X_{J}]^{2}\Bigr)  + \nonumber \\ 
&& \mbox{fermionic term}.  
\end{eqnarray} 
Here the indices $i,j,k,l$ run from $1$ to $d$, the indices $I,J$ run from $d+1$ to $9$, $\nabla_{i} = \nabla_{i}^{0} + iA_{i}$ 
stand for the Yang-Mills fields and $X_{I}$ for the scalar fields, $G^{ij}$ is the torus metric and $g_{YM}$ is the Yang-Mills 
coupling constant.

To describe a D0-D2 system on $T_{\theta}^{2}$ we consider the action functional (\ref{action}) for $d=2$ with the 
fields acting on a module $E_{1,1}\cong E_{0,1}\oplus E_{1, 0}$ described in the previous section. 
There exists a $\frac{1}{2}$-BPS solution  on this module describing a D0 brane dissolved inside a D2 brane wrapped on $T_{\theta}^{2}$. 
 Our prime interest here is however in an unstable soliton describing a point-like D0-brane  sitting on top of D2 brane 
(or more generally being displaced away in the transverse  to  the D2-brane direction by some small distance). 
This soliton is a torus analog of 
the solitonic solution of \cite{unstable} on a noncommutative two-plane. Below we will refer to the unstable configuration as 
an undissolved D0 on D2 system.  
 The BPS configuration describes a bound state that should be  the  end
 point of tachyon condensation in the unstable  system. We will have a further discussion of the BPS state later when 
computing the soliton binding energy.

We  propose to use the following static solution of SYM equations 
to describe the  undissolved D0 on   D2  system on $T_{\theta}^{2}$
\begin{eqnarray} \label{sol}
\nabla_{j} = \left( 
\begin{array}{cc}
\tilde \nabla^{0}_{j}  + ic_{j}\cdot{\bf 1}& 0 \\
0 & \partial_{j}   
\end{array} \right) \, , \quad X^{I} = 
\left( 
\begin{array}{cc}
 x^{I}\cdot{\bf 1}& 0 \\
0 & 0
\end{array} \right) \, . 
\end{eqnarray}
Here the constants $c_{j}$ describe the position of D0 brane on a torus and $x^{I}$'s are its transverse coordinates 
(see a more detailed discussion later). 

The curvature of this solution has a block form 
$$
F_{12} = \left( 
\begin{array}{cc}
 -\frac{2\pi i }{\theta}{\bf 1} & 0 \\
0 & 0
\end{array} \right)
$$
and the energy is 
\begin{equation} \label{sol_energy}
E_{sol} = \frac{(2\pi)^{2}}{4 g_{YM}^{2}}\sqrt{{\rm det}(G\theta )} \left(\frac{1}{\theta}\right)_{ij}G^{ik}G^{jl} 
\left(\frac{1}{\theta}\right)_{lk}  \, . 
\end{equation}
In two dimensions this expression reduces to 
\begin{equation} \label{2d_en}
\frac{(2\pi)^{2}}{2g_{YM}^{2}\sqrt{{\rm det} G} \theta} \, .
\end{equation}

\noindent \underline{\bf Soliton energy vs. D0-D2 binding energy.}
The excess of the soliton energy (\ref{2d_en}) above the Yang-Mills energy of the bound state should be 
compared to the binding energy of the D0-D2 system in a constant $B$-field. Note that on the plane 
the process of tachyon condensation goes through a spreading of the D0-brane flux over the infinite volume of the plane. 
The Yang-Mills energy of the limiting bound state vanishes and can be neglected in the computation. This is not 
the case for the torus. The  BPS solution (the explicit form of which will be discussed in the last section) has 
a constant curvature that in the block decomposition form can be written as 
$$
F_{12} = \left( 
\begin{array}{cc}
 -\frac{2\pi i }{\theta + 1}{\bf 1} & 0 \\
0 &   -\frac{2\pi i }{\theta + 1}{\bf 1}
\end{array} \right) \, . 
$$

Its energy is 
$$
E_{BPS}^{SYM} = \frac{(2\pi)^{2}}{2g_{YM}^{2}\sqrt{{\rm det} G} (\theta + 1)} 
$$
and the soliton excess energy is 
\begin{equation} \label{deltaE}
\Delta E = E_{sol} - E_{BPS}^{SYM} = \frac{(2\pi)^{2}}{2g_{YM}^{2}\sqrt{{\rm det} G} \theta (\theta + 1)} \, . 
\end{equation}

To compute the D0-D2 binding energy in the presence of $B$-field we can use the Born-Infeld energy. 
In the Seiberg-Witten decoupling limit 
\begin{equation} \label{SW_lim}
\alpha'\sim \epsilon^{1/2}\to 0 \, , \qquad g_{ij}\sim \epsilon\to 0 
\end{equation} 
the energy of the D2-brane has an expansion 
$$
E_{D2} = T_{2}\sqrt{ {\rm det}g + {\rm det}(2\pi \alpha'B)} = T_{2}2\pi \alpha' B + \left(\frac{T_{2}}{2}\right)
\frac{ {\rm det}g}{2\pi \alpha' B} + 
O(\epsilon)  
$$
where $T_{2} = (g_{s}(2\pi)^{2}(\alpha')^{3/2})^{-1}$ is the 2-brane tension. 
The energy of the bound state can be evaluated from the above Born-Infeld by adding the corresponding  constant field strength 
as 
$$
E_{D2/D0} = T_{2}\sqrt{ {\rm det}g + {\rm det}(2\pi \alpha'(B + 2\pi) )} = 
T_{2}2\pi \alpha' B + \left(\frac{T_{2}}{2}\right)\frac{ {\rm det}g}{2\pi \alpha' (B + 2\pi)} + 
O(\epsilon) \, . 
$$
Thus  the D0-D2 binding  energy is  
\begin{equation} \label{E_bin2D}
 E_{bind}  = T_{0} + E_{D2} - E_{D2/D0} = \frac{T_{2} \, {\rm det} g}{2\alpha' B(B+2\pi)} \, . 
\end{equation}

To compare  this expression with the soliton excess energy  we  rewrite (\ref{deltaE}) in terms of closed string 
parameters $B$ and $g$. 
It follows from our periodicity conventions $x^{i}\sim x^{i} + 1$ that 
$2\pi/\theta = B$. In the Seiberg-Witten limit (\ref{SW_lim}) on a D2p-brane  
we have in the leading order \cite{SW}  
\begin{eqnarray} \label{cl_par}
&& G_{ij} = -(2\pi \alpha')^{2}(Bg^{-1}B)_{ij} \, , \nonumber \\ 
&&  \frac{1}{g_{YM}^{2}}= 
\frac{\sqrt{{\rm det}g}}{|{\rm Pf}B| g_{s} (\alpha')^{2p-3/2}(2\pi)^{3p-2}} = 
\frac{\sqrt{{\rm det}g} T_{2p}}{|{\rm Pf}B|(2\pi)^{p-2}(\alpha')^{p-2}}\, .  
\end{eqnarray}
Taking $p=1$ and substituting these expressions in (\ref{deltaE}) we find an exact agreement with  
(\ref{E_bin2D}). 

 This result  supports our proposal 
that the solution (\ref{sol}) describes the undissolved D0 on  D2 system on a torus with a uniform magnetic field turned on. 
A further support comes from matching 
 the spectrum of quadratic fluctuations around  (\ref{sol})  with that of the D0-D2 excitations in the  
decoupling limit (\ref{SW_lim}) taken on a torus. 


\noindent \underline{\bf Fluctuations.}
Let us represent fluctuations around  (\ref{sol}) in the form 
$$
\delta A_{j} = \left( 
\begin{array}{cc}
 \delta z_{j}& \delta\psi_{j} \\
\delta \psi_{j}^{*} & \delta u_{j}
\end{array} \right) \, , \quad 
\delta X^{I} = \left( 
\begin{array}{cc}
 \delta a^{I}& \delta \chi^{I} \\
  \delta \chi^{*I} & \delta b^{I}
\end{array} \right) \, . 
$$
The Gauss constraint 
$$
[\nabla_{i}, \partial_{t}A^{i}] + [X_{I}, \partial_{t}X^{I}] = 0
$$ 
being expanded to the first order in fluctuations gives 
\begin{eqnarray} \label{Gauss}
&& \partial_{t}([\tilde \nabla_{i}^{0}, \delta z^{i}]) = 0 \, , \qquad \partial_{t}([\partial_{i}, \delta u^{i}]) = 0 \, , 
\nonumber \\
&& \partial_{t}( \tilde \nabla_{i}^{0}\delta \psi^{i} + ic_{i}\delta \psi^{i} + x_{I}\delta\chi^{I} ) = 0 \, .
\end{eqnarray}

The calculation of the second derivative matrix is  similar to the analogous computation on the plane. There is 
only one point in the computation that we would like to highlight. When encountering expressions of the form 
${\rm Tr}_{1/\theta}\delta \psi_{i} \circ \delta\psi^{*}_{j}$ or ${\rm Tr}_{\theta}\delta \chi^{*}_{i} \circ \delta\chi_{j}$ 
we use the identity (\ref{identity}) that gives us $\langle \delta \psi_{j}, \delta \psi_{i} \rangle_{\mathbb C}$ and  
 $\langle \delta \chi_{i}, \delta \chi_{j} \rangle_{\mathbb C}$ respectively. With this in mind it is straightforward to obtain 
$$
-\frac{1}{2}\delta^{2}S = \frac{\sqrt{G}}{g_{YM}^{2}}( K_{2-2} + K_{0-0} + K_{0-2}) 
$$
where 
\begin{equation} \label{K2}
K_{2-2}= {\rm Tr}_{\theta}\Bigl( \frac{1}{2}(\partial_{t} \delta u_{i})^{2}   +  \frac{1}{2}(\partial_{t} \delta b^{I})^{2} +
\frac{1}{4} ([\partial_{i}, \delta u_{j}] - [\partial_{i}, \delta u_{j}])^{2}  + \frac{1}{2}[\partial_{i}, \delta b^{I}]^{2}      \Bigr) 
\end{equation}
\begin{equation} \label{K0}
K_{0-0}= {\rm Tr}_{1/\theta}\Bigl( \frac{1}{2}(\partial_{t} \delta z_{i})^{2}   +  \frac{1}{2}(\partial_{t} \delta a^{I})^{2} +
\frac{1}{4}([\tilde \nabla_{i}^{0}, \delta z_{j}] - [\tilde \nabla_{j}^{0}, \delta z_{i}])^{2}  + 
\frac{1}{2}[\tilde \nabla_{i}^{0}, \delta a^{I}]^{2}      \Bigr) 
\end{equation}
\begin{eqnarray} \label{K0-2}
&& K_{0-2} = \| \partial_{t}\delta \psi^{i}\|^{2}_{\mathbb C} +  
\|\partial_{t}\delta \chi^{I}\|^{2}_{\mathbb C} + 
\frac{1}{2}\| \tilde \nabla_{i}^{0}\delta \psi_{j} - 
\tilde \nabla_{j}\delta \psi_{i} +i(c_{i}\delta \psi_{j}-c_{j}\delta \psi_{i})\|^{2}_{\mathbb C} 
- \nonumber \\ 
&& ({\rm det}G)^{-1} \frac{2\pi i}{\theta} 
(\langle \delta \psi_{2}, \delta \psi_{1}\rangle_{\mathbb C} - \langle \delta \psi_{1}, \delta \psi_{2}\rangle_{\mathbb C}) + 
 \|\tilde \nabla_{i}^{0}\delta \chi^{I} + ic_{i}\delta \chi^{I} -ix^{I}\delta\psi_{i}\|^{2}_{\mathbb C}
\end{eqnarray}
where all of the summations in indices $i, j$ are performed with the metric tensor $G^{ij}$.

It is not hard to see from the expression for $K_{0-2}$ 
that despite a more complicated structure (\ref{inner_pr}) of the off-diagonal terms, in the matrix of quadratic 
fluctuations  their contribution  is exactly the same as on the plane 
\cite{unstable}, \cite{ohta}. This happens  due to  identities  (\ref{identity}). 
This of course goes along well  with our physical interpretation because  on the torus the 0-2 
strings have the same spectrum as   on the plane. 
Our  analysis of  (\ref{K0-2}) is essentially the same as in  \cite{unstable}. 
In order to have a complete discussion and bearing in mind the difference in notation that might obscure the correspondence with 
\cite{unstable} we would like to sketch here the analysis of $K_{0,2}$. 
For simplicity let us first consider the case $x^{I}=0$ when D0 seats on top of D2. Let us also 
assume here   $G_{ij} = G\delta_{ij}$ (we will restore  a generic constant metric  later in the analysis of diagonal fluctuations 
where it will be truly important). 
Recall that in the Fock space picture of the module $E_{0,1}$ the operators $\tilde \nabla_{j}^{0}$ are expressed 
via creation and annihilation operators according to  (\ref{n-a}). 
It is easy to see that the spectrum of fluctuations does not depend on the moduli $c_{j}$ as  one can absorb them into 
 redefined creation and annihilation operators related to the ones introduced in (\ref{aa}) by a suitable unitary transformation on 
$L_{2}({\mathbb R}^{1})$.  
 Changing the variables 
$$
\delta\psi_{1} = i(\delta \psi_{-} -\delta\psi_{+})\frac{1}{\sqrt{2}}\, , \qquad 
\delta\psi_{2} = (\delta \psi_{-} +\delta\psi_{+})\frac{1}{\sqrt{2}}
$$  
and using (\ref{n-a}) we obtain for the potential term in (\ref{K0-2})
$$
\frac{2\pi }{\theta G^{2}}( \|a\delta \psi_{-} + a^{\dagger}\delta \psi_{+}\|^{2} + \|\delta \psi_{+}\|^{2} - \|\delta \psi_{-}\|^{2} + 
G \langle \delta \chi^{I}, (1 + 2a^{\dagger}a)\delta \chi^{I}\rangle_{\mathbb C}  ) \, . 
$$  
Expanding 
$$
\delta \psi_{\pm} = \sum_{n\ge 0} t_{n}^{\pm } |n\rangle \, , \quad \delta \chi^{I} =  \sum_{n\ge 0} s_{n}^{I}|n\rangle
$$ 
and substituting into the above expression we obtain 
\begin{eqnarray*}
K_{0-2} &=& \sum_{n=0}^{\infty}\Bigl( \frac{1}{2G} | \partial_{t}t_{n}^{+}|^{2} +   
\frac{1}{2G} | \partial_{t}t_{n}^{-}|^{2} + 
\frac{1}{2} |\partial_{t}s_{n}^{I}|^{2}  + \nonumber \\
&& \frac{2\pi }{\theta G^{2}}\Bigl( \sum_{n=0}^{\infty} |\sqrt{n}t^{+}_{n-1} + \sqrt{n+1}\psi^{-}_{n+1}|^{2} +
|t^{+}_{n}|^{2} - |t^{-}_{n}|^{2} + G(2n+1) |s_{n}^{I}|^{2} 
\Bigr)
\end{eqnarray*}
It follows from this expression that the bosonic spectrum of 0-2 strings contains a tachyon mode $t_{0}^{-}$ 
with mass squared $m_{tach}^{2}=-2\pi /(G \theta )  = -g(2\pi \alpha')^{-2}B^{-1}$, a tower of positive mass modes 
$$
\xi_{n} = \frac{1}{\sqrt{2n +1}}(\sqrt{n+1}\psi^{+}_{n-1} + \sqrt{n}\psi^{-}_{n+1})
$$  with 
$m^{2} = (2n+1)g(2\pi \alpha')^{-2}B^{-1}$, 
 the same tower for each transverse scalar $\chi^{I}$ and an infinite number of zero modes 
$$
\eta_{n} = \frac{1}{\sqrt{2n+1}}(\sqrt{n}\psi_{n-1}^{+} - \sqrt{n+1}\psi^{-}_{n+1})
$$  
that are nondynamical due to the Gauss law constraint (\ref{Gauss}). 
The analysis of 0-2 strings in the Seiberg-Witten decoupling limit is exactly the same on the torus as on the plane, 
thus as it was shown in \cite{unstable} we have a match of the  fluctuation spectra in this sector. 
It is easy to see from (\ref{K0-2}) that switching on the moduli  $x^{I}$ shifts all squared masses up by 
$x^{I}x_{I} $. This matches with the shift in the spectra of D0-D2 system that occurs when we displace D0 in a 
 direction transverse to the torus by the vector $2\pi \alpha' x^{I}$.

Let us discuss next the spectrum of fluctuations entering the $K_{0-0}$ term. Expanding 
$$
\delta z_{j} = \sum_{{\bf n}\in {\mathbb Z}^{2}} q_{j}({\bf n}) Z_{\bf n} \, , \quad \delta 
a^{I} = \sum_{{\bf n}\in {\mathbb Z}^{2}} r^{I}({\bf n}) Z_{\bf n}
$$
imposing the Gauss law constraint (\ref{Gauss}) 
and using the formula 
\begin{equation} \label{zn}
[\tilde \nabla_{j}^{0}, Z_{\bf n}] = \frac{2\pi i}{\theta}   n_{j} Z_{\bf n} \, ,   
\end{equation}
that  follows from (\ref{Zi}), (\ref{ccc}), 
 we obtain that the fluctuations are described in terms of a set of 8 towers of  oscillators with frequencies   
\begin{equation} \label{omega}
\omega^{2} ({\bf n}) = \left(\frac{2\pi}{\theta}\right)^{2} n_{i}G^{ij}n_{j} 
\end{equation}  and  polarization transverse to ${\bf n}$.
On the other hand, for a D0-brane on a torus with closed string metric $g_{ij}$ and radii $r=1/2\pi$ 
the winding modes $w^{i}$ contribute to the spectrum as 
$$
m^{2} = \left(\frac{1}{2\pi \alpha'}\right)^{2}w^{i}g_{ij}w^{j} + \frac{1}{\alpha'}(N - a) \, . 
$$
In the Seiberg-Witten limit the surviving states in the bosonic sector are $e_{m}\psi_{-1/2}^{m}|0, w_{1}, w_{2}\rangle_{NS}$ 
where $e_{m}$ is the polarization vector for the collective coordinates of D0-brane.  
We have therefore 8 transverse oscillators with  
$$
m^{2} = -w^{i}(BGB)_{ij}w^{j} = \frac{1}{\theta^{2}} w^{i}\epsilon_{ij}G^{jk}\epsilon_{kl}w^{l} 
$$
that matches with (\ref{omega}) upon identification $n_{i} = \epsilon_{ij}w^{j}$.

The term $K_{2-2}$ similarly gives us a tower of K-K modes of noncommutative Yang-Mills theory with 7 adjoint scalars 
on a torus that is 
precisely what we expect for excitations of a wrapped D2-brane. Thus we see that the spectra in all sectors match. 


\noindent \underline{\bf Moduli.} 
Now we would like to discuss the moduli $c_{i}$. It is clear that these moduli should come from  the positions 
of D0-branes on the original shrinking torus. The analogous moduli and their interpretation for the case of noncommutative 
plane were discussed in \cite{unstable}, \cite{GrossNek}, \cite{Rangamani}, \cite{ohta}, \cite{Hamanaka} and more recently in \cite{ohta2}.
On the torus we need to check the periodicities of $c_{j}$ and a metric on them. 
 We first note that due to gauge 
transformations 
$$
Z_{\bf n}^{\dagger}(\tilde \nabla_{j}^{0} + ic_{j}{\bf 1})Z_{\bf n} = \tilde \nabla_{j}^{0} + i(c_{j} + \frac{2\pi}{\theta}
 n_{j}){\bf 1} 
$$
the moduli $c_{i}$ are periodic with periods $\frac{2\pi}{\theta}$.  A natural  metric on moduli $c_{i}$ is $G^{ij}$. 
Probably the most convincing way to check the last statement is to consider a module representing two D0 branes 
on top of a D2-brane (more on these generalized solutions in the next section). 
It is given by a direct sum of two copies of D0-module $E_{0,1}$ and a single copy of a D2-module 
$E_{1,0}$. Then the appropriate soliton has a form (\ref{sol}) with an upper block replaced by a two-by-two block 
$$
\left( 
\begin{array}{cc}
\tilde \nabla^{0}_{j}  + ic_{j}^{(1)}{\bf 1}& 0 \\
0 & \tilde \nabla^{0}_{j}  + ic_{j}^{(2)}{\bf 1}  
\end{array} \right) \, . 
$$  
It can be checked then that the off-diagonal fluctuations in the above block are massive with masses 
$$
m^{2}= (c_{i}^{(1)}-c_{i}^{(2)})G^{ij}(c_{j}^{(1)}-c_{j}^{(2)}) \, . 
$$ 
If we introduce now rescaled coordinates $\tilde c_{j} = \frac{\theta}{2\pi}c_{j}$ that have periodicities 1, 
then the metric on these coordinates is $\frac{2\pi}{\theta}G^{ij}$. On the other hand for the
strings 
stretched between two D0 branes on a torus located at points $x_{1}^{i}$ and $x_{2}^{i}$ we have the mass spectrum 
of the form 
$$
m^{2} = \left(\frac{2\pi}{\alpha'}\right)^{2}(x_{1}^{i} - x_{2}^{i})g_{ij}(x_{1}^{j} - x_{2}^{j}) + \frac{1}{\alpha'}(N-a) 
$$ 
that in the Seiberg-Witten limit  gives  the metric  $-\frac{2\pi}{\theta}\epsilon_{ik}G^{kl}\epsilon_{lj}$. 
Thus we can identify 
\begin{equation}\label{mod}
c_{i}^{a} = \frac{2\pi}{\theta}\epsilon_{ij}x_{a}^{j}\, . 
\end{equation}

Note that we could also switch on Wilson lines on the D2-brane by taking 
$\partial_{j} + is_{j} {\bf 1}$ instead of $\partial_{j}$ in (\ref{sol}). 
In that case a gauge transformation 
$$
U^{\dagger}_{\bf n}(\partial_{j} + is_{j} {\bf 1})U_{\bf n} = \partial_{j} + i(s_{j} + 2\pi n_{j}){\bf 1}
$$ 
implies that the moduli $s_{j}$ have periodicities $2\pi$. A natural metric on them is $G^{ij}$. 
It can be checked that in the Seiberg-Witten limit 
$$
\left(\frac{2\pi}{\alpha'}\right)^{2} g'_{ij} \to -\frac{1}{(\alpha')^{2}}\left( \frac{1}{B}g\frac{1}{B}\right)^{ij} = 
       (2\pi)^{2} G^{ij}
$$
where 
$$
g'_{ij}= \left( \frac{\alpha'}{g-2\pi\alpha' B}\, g\, \frac{\alpha'}{g+2\pi\alpha' B}\right)_{ij}
$$ is a metric obtained from $g_{ij}$, $B_{ij}$ by a T-duality in all sides of the torus. Thus 
the moduli $s_{j}$ specify a point on the T-dual torus as it should be in the usual interpretation of Wilson lines.

Finally we would like to remark that the most direct way to see the localization of D0-brane and to check the metric on its positions 
moduli would be through the use of Seiberg-Witten transformation. Recently the Seiberg-Witten transform of noncommutative solitons and 
instantons on a plane was studied in papers \cite{HO}, \cite{KS}. 


\section{Generalizations} \label{h.d.}
In this section we will consider both generalizations for higher dimensional tori and 
for systems with multiple   numbers of branes.  

We start with a solution describing an undissolved D0-brane seating on top of D2p brane wrapped on $T^{2p}$.
For simplicity we will assume that our noncommutativity matrix is brought to the canonical form 
\begin{equation}\label{th}
(\theta^{ij}) = \left( \begin{array}{cccc} 
\theta_{1}{\bf \epsilon} & 0 & \dots & 0\\
0 & \theta_{2}{\bf \epsilon} & \dots & 0\\
\vdots  & \vdots & \ddots & \vdots\\
0 & 0 & \dots  & \theta_{p}{\bf \epsilon} 
\end{array} \right)  
\end{equation}
where 
$$
{\bf \epsilon} = \left( \begin{array}{cc}
0&1\\
-1&0
\end{array} \right) \, . 
$$

A D2p-module is again just a rank one free module over  $T_{\theta}^{2p}$. 
A D0-module is constructed by defining the action of torus generators on functions of $p$-variables 
$x^{1}, \dots , x^{p}$ as  
\begin{eqnarray} \label{D0}
\phi(x)U_{2j + 1}= \phi(x^{1}, \dots ,x^{j}-\theta_{j} , \dots x^{p}) \, , \qquad 
\phi(x)U_{2j}=e^{2\pi i \theta_{j}x^{j}}\phi(x) \, , \enspace j=1, \dots , p \, . 
\end{eqnarray}
This module has a constant curvature connection 
\begin{equation} \label{ccc2}
\tilde \nabla_{2j + 1}^{0} \phi(x)= \frac{2\pi i }{\theta_{j}}x^{j}\cdot \phi(x) \, , \qquad 
\tilde \nabla_{2j}^{0} \phi(x)= \partial_{j}\phi(x)  
\end{equation} 
with the curvature tensor  
$$
[\tilde \nabla_{j}^{0}, \tilde \nabla_{k}^{0}]= 2\pi i (\theta^{-1})_{jk}{\bf 1} \, . 
$$

By computing the Chern character and then using Elliott's formula \cite{Elliott} (see e.g. \cite{review1} for an 
explanation of these tools in noncommutative geometry) one can show that this module has a D0 number equal to 1 and no other 
charges \cite{moduli}. 
Next we take a module  that is a direct sum of the D0-module (\ref{D0}) and a D2p-module  
 and consider a SYM solution that upon using the connection (\ref{ccc2})  
for $\tilde \nabla_{j}^{0}$  has exactly the same  form as (\ref{sol}). 
Its energy is given by (\ref{sol_energy}) where $G$ and $\theta$ are now understood as $2p\times 2p$ matrices.

\noindent \underline{\bf Soliton energy and D0-D4 binding energy.}
Next we would like to compute the soliton excess energy above the BPS state for $p=2$ and compare it to the binding energy 
of the D0-D4 system. The BPS solution is a noncommutative instanton. It preserves $\frac{1}{4}$ of  supersymmetries. 
Its energy can be computed from the noncommutative SYM supersymmetry algebra and reads \cite{susyalg} 
\begin{equation} \label{4dBPS}
E^{SYM}_{BPS} = \frac{ \sqrt{{\rm det}G} \pi^{2}} {g_{YM}^{2}(1 + {\rm Pf}\theta )} |{\rm tr}(\tilde \theta G^{-1}\tilde \theta 
G^{-1})| + \frac{(2\pi)^{2}} {g_{YM}^{2}(1 + {\rm Pf}\theta)} 
\end{equation} 
where $\tilde \theta_{ij} = \frac{1}{2}\epsilon_{ijkl}\theta^{kl}$ and ${\rm Pf}\theta$ is assumed 
to be positive. 
Let us explain here  where this choice comes from. For a general module $E$ over $T_{\theta}^{4}$ that 
carries $n$ D4, $m$ D0 and no other charges the noncommutative dimension is ${\rm dim}(E) = n + m{\rm Pf}\theta$ 
and it has to be positive, otherwise this charges specify just a virtual module (a representative 
of the $K_{0}$ group). When $n=1$ and $m=1$ we can either have  ${\rm Pf}\theta>0$ or $ -1<{\rm Pf}\theta<0$. 
In the last case however the module cannot be represented as a direct sum of a D4 and a D0 module, but rather 
as a difference (in the sense of K-theory). Thus to have simultaneously both the undissolved and the dissolved BPS
solutions on our module we should take   ${\rm Pf}\theta>0$. This choice also follows from the analysis 
of the binding energy made below. The system with  $ -1<{\rm Pf}\theta<0$ behaves rather like a  $\overline {\rm D0}$-D4 
system (see more on this at the end of this section). 
The qualitatively different  behavior for  different signs of ${\rm Pf}B$ in the D0-D4 system is discussed in 
\cite{SW}, \cite{David}.  Note that in this paper we have the opposite sign convention  than in those two papers 
(this comes from the use of Elliott's formula (\ref{Elliott})).

The soliton excess energy above the BPS state can be computed from (\ref{sol_energy}), (\ref{4dBPS}) to be 
\begin{equation} \label{dE}
\Delta E = \frac{ \sqrt{{\rm det}G} \pi^{2}} {g_{YM}^{2}{\rm Pf} \theta (1 + {\rm Pf}\theta )} |{\rm tr}(\tilde \theta G^{-1}\tilde \theta 
G^{-1})| - \frac{(2\pi)^{2}}{g_{YM}^{2}(1 + {\rm Pf}\theta)}
\end{equation}
where we used the identity 
$$
{\rm tr}\, \frac{1}{\theta}G^{-1}\frac{1}{\theta}G^{-1} = 
\frac{ {\rm tr}\, \tilde \theta G^{-1}\tilde \theta G^{-1}} {({\rm Pf} \theta)^{2}} \, . 
$$

The mass squared for a $\frac{1}{4}$ BPS state with $N_{4}$ D4-branes and $N_{0}$ D0-branes in the presence of $B$-field 
is given by the following expression taken from \cite{OP} (see their formula (3.42)) 
\begin{eqnarray} \label{OPf}
M^{2} & =& T_{4}^{2}\Bigl[ N_{4}^{2}{\rm det}g + (2\pi \alpha')^{4}(N_{4}{\rm Pf}B + (2\pi)^{2}N_{0})^{2} + 
\nonumber \\
&& \frac{1}{2}N_{4}^{2}(2\pi\alpha')^{2}|{\rm tr}\, \tilde Bg\tilde Bg| + 
2(2\pi\alpha')^{2}(2\pi)^{2}|N_{0}N_{4}|\sqrt{{\rm det}g}\Bigr] \, . 
\end{eqnarray} 
For $N_{0}=1$, $N_{4}=1$ we have in the limit (\ref{SW_lim}) 
\begin{equation}
E_{D4/D0}=T_{4}(2\pi\alpha')^{2}({\rm Pf}B + (2\pi)^{2}) + \frac{T_{4}}{({\rm Pf}B + (2\pi)^{2})}
[\frac{1}{4}|{\rm tr}\, \tilde Bg\tilde Bg| + (2\pi)^{2}\sqrt{{\rm det} g}] + O(\epsilon) \, . 
\end{equation}
On the other hand for the energy of a single D4-brane in a $B$-field we have from  (\ref{OPf} (or equivalently from 
Born-Infeld action) 
\begin{equation}
E_{D4} = T_{4}(2\pi\alpha')^{2}{\rm Pf}B + \frac{T_{4}|{\rm tr}\, \tilde Bg\tilde Bg|}{4{\rm Pf}B} + O(\epsilon) \, . 
\end{equation}
Taking the difference we obtain for the binding energy in the leading order 
\begin{equation}\label{4d_bin}
E_{bind} = T_{0} + E_{D4} - E_{D4/D0} = \frac{(2\pi)^{2}T_{4} }{ ({\rm Pf}B + (2\pi)^{2})}\Bigl[ 
\frac{|{\rm tr}\, \tilde Bg\tilde Bg|} {4{\rm Pf}B} - \sqrt{{\rm det}g}\Bigr] \, . 
\end{equation}

Upon expressing (\ref{dE}) in terms of closed string parameters (\ref{cl_par}) we find  again an exact agreement with 
the binding energy (\ref{4d_bin}). When checking this we found the following identity useful 
$$
{\rm tr} \, \tilde Bg\tilde Bg = {\rm det}g\, {\rm tr}\, Bg^{-1}Bg^{-1} \, . 
$$

The analysis of fluctuations around the $2p$-dimensional soliton is  parallel to the analysis 
for the two-dimensional case done in the previous section. In fact for the 0-0 and 4-4 strings the results for $2p=2$ 
can be generalized in a  straightforward   manner. As for the 0-4 strings it is simple to show using (\ref{identity}) 
that the quadratic fluctuation matrix in the end reduces to the one on the plane. We can use then the results of 
paper (\cite{ohta}) in which the agreement of the soliton fluctuation spectrum  and that of the 0-2p string excitations was 
checked for $p=1,2,3,4$. 

Furthermore, in analogy with formulae (\ref{Zi}) and (\ref{zn}) 
we can introduce a basis $Z_{\bf n}$ for the endomorphisms  (generators of the 0-0 strings algebra) 
 such that 
$$
[\nabla_{j}, Z_{\bf n}] = 2\pi i \theta^{-1}_{jk}n^{k}  Z_{\bf n} \, . 
$$   
Then it is not hard to see that our discussion of the moduli generalizes to the case at hand and yields  
an  identification  
$$
c_{j} = 2\pi \theta^{-1}_{jk}x^{k}
$$
analogous to (\ref{mod}). A similar statement  is also true for  the Wilson lines moduli on D2p.

\noindent \underline{\bf Branes of intermediate dimension.} 
So far we were discussing systems that contained branes of extremal dimensions. And these type of systems 
are  in some sense direct descendants  of the similar systems on a plane (sphere). It would be more interesting to 
see solitons describing undissolved  branes of intermediate dimension. Here we would like to propose a module and a soliton 
that we conjecture describes a two-brane on top of a four-brane on $T_{\theta}^{4}$. It will be clear from this particular 
example how one can generalize this solution to other systems with intermediate branes.

First note that with the choice of $\theta^{ij}$ in the form (\ref{th}) the algebra of functions on a four-dimensional 
noncommutative torus $T_{\theta}^{4}$ is actually a tensor product $T_{\theta_{1}}^{2}\otimes T_{\theta_{2}}^{2}$ 
of two mutually commuting algebras of functions on two-dimensional tori. Thus we can consider modules  
that are tensor products $F\otimes G$. Here $F$ is a module over   $T_{\theta_{1}}^{2}$ and $G$ is a module over 
$T_{\theta_{2}}^{2}$. Now take $F=E_{0,1}$, $G=E_{1,0}$ where $E_{0,1}$ and $E_{1,0}$ are modules introduced in section 2. 
The Chern character for $E=F\otimes G$ is a product of the Chern characters of the factors: 
${\rm ch}(E) = {\rm ch}(F) {\rm ch}(G)= 1\cdot (1 + \alpha^{2}\wedge \alpha^{3})= 1 + \alpha^{2}\alpha^{3}$ 
where $\alpha^{i}$ stand for the noncommutative analogs of the integer cohomology generators.
Thus we see that $E$ carries charges corresponding to a D2-brane wrapped on the (1,2)-cycle. 
Furthermore this module has a connection 
$$
\tilde \nabla_{i} = \tilde \nabla^{0}_{i}\otimes {\bf 1} \, , \enspace \mbox{for}\enspace i=1,2 \qquad 
\tilde \nabla_{i} = {\bf 1}\otimes \partial_{i} \, , \enspace \mbox{for}\enspace i=3,4 
$$
where $\tilde \nabla^{0}_{i}$ is given by (\ref{ccc}). This connection has  a curvature tensor whose only nonvanishing 
 components are
$F_{12} = -F_{21}=-\frac{2\pi i}{\theta_{1}}\cdot 1$. To describe a D2-D4 system let us take a direct sum 
$E=E_{1}\oplus E_{2}$ 
of a module we have just described, to be denoted $E_{1}$, and a rank one free module $E_{2}=(T_{\theta}^{4})^{1}$. 
Let us take a Yang-Mills field  $\nabla_{i}$ to be  a direct sum of $\tilde \nabla_{i}$ described above and a 
standard connection $\partial_{i}$ on $E_{2}$. Then it is straightforward to see that we have a solution to the Yang-Mills 
equations of motion (for simplicity we set here to zero the analogs of  moduli $c_{i}$ and $x^{I}$). 
We also know that  this module carries the correct D2 and D4 charges. Furthermore it is not hard to see using the 
results we already obtained that we have the correct spectrum of fluctuations. Let us sketch how it works. 
Assume for simplicity that the metric $G^{ij}$ also has a block-diagonal  form with diagonal 
blocks giving  a metric on each  $T^{2}$.
Fluctuations around the soliton have schematically the following form 
$$
\delta A_{j}= \left( \begin{array}{cc} 
 \delta u_{j}\otimes \delta z_{j} &  \delta u'_{j}\otimes\delta \psi_{j}\\
(\delta u_{j}')^{*}\otimes\delta \psi_{j}^{*} & \delta u_{j}'' 
\end{array} \right)
$$
where $\delta z_{j}:F\to F$ are endomorphisms of $F$ (if we forget about the second torus these are 
elements of the 0-0 strings  algebra on $T_{\theta_{2}}^{2}$), 
$\delta \psi_{j}$ are elements of  $F$ itself, $\delta u_{j}, \delta u_{j}'\in T_{\theta_{2}}^{2}$, 
$\delta u_{j}''\in T_{\theta}^{4}$. The string spectrum in the D2-D4 system is also built as a tensor product 
of spectra for individual directions $x^{1}, x^{2}, x^{3}, x^{4}$. It is clear then that 
we should identify the fluctuations $\delta z_{j}$ and $\delta u_{j}$ with those of D-D and N-N  directions  
in the 2-2 string excitations, $\delta \psi_{j}$ and $\delta u_{j}'$ with D-N and N-N directions for 2-4 
strings, and $\delta u_{j}''$ with excitations of 4-4 strings. Indeed, take for example the 2-2 sector. 
We showed already that the modes $\delta z_{j}$ on $T_{\theta_{2}}^{2}$ give the correct spectrum of D0-D0 winding 
modes. And this is precisely what we should have for the two D-D directions of the 2-2 strings as they can wind around 
the second torus. It is easy to see that the other sectors match as well.

\noindent \underline{\bf Multiple branes.} 
Once we constructed individual D2p-brane modules it is simple to construct modules describing 
multiple brane systems. To that end one simply takes direct sums of modules and elementary solutions 
on each one of them.  

For example if we would like to describe  a system of  $m$ D0-branes sitting on top of a stack of  $n$ D2 branes 
wrapped on a two-torus  we  take a direct sum of $m$ copies of D0 module and $n$ copies of the D2 module.  
The four blocks in (\ref{sol}) get enlarged to  $m\times m$, $m\times n$ and $n\times n$ matrices as 
\begin{equation} \label{nmsol}
\nabla_{j} = \left( 
\begin{array}{cc}
\tilde \nabla_{j}\otimes {\bf 1}_{m\times m}  + ic_{j}& 0 \\
0 & \partial_{j}\otimes  {\bf 1}_{n\times n} 
\end{array} \right) \, , \quad X^{I} = 
\left( 
\begin{array}{cc}
 x^{I} & 0 \\
0 & 0
\end{array} \right) \, . 
\end{equation}
where $c_{j}$ and $x^{I}$ are now $m\times m$ diagonal matrices whose eigenvalues describe positions of $m$ D0 branes.
The analysis of fluctuations 
for this and analogous systems is a direct generalization of the one we described in  section 2.
 Fluctuations in the upper and lower diagonal 
blocks in  (\ref{nmsol} are those of noncommutative Yang-Mills theories with gauge groups $U(m)$ and $U(n)$ respectively and  with 
gauge fields taking values  in the appropriate algebras:  in the $T_{\theta}$ and  in the T-dual torus $T_{1/\theta}$. 
They describe the corresponding sectors of various D0-D0 and D2-D2 strings. The off diagonal terms correspond to the $n\times m$ 
sectors of D0-D2 strings each one described by a vector $\delta \psi_{ij} \in L_{2}({\mathbb R}^{1})$.

 A general projective module $E$ over $T_{\theta}^{2p}$ is described by a collection of charges that can be 
written out as an inhomogeneous exterior form with integer coefficients 
$$
\mu(E) = n + \frac{1}{2}m_{ij}\alpha^{i}\wedge \alpha^{j} + \frac{1}{4!}m^{ijkl}
\alpha^{i}\wedge \alpha^{j}\wedge \alpha^{k}\wedge \alpha^{l} + \dots 
+ m\alpha^{1}\wedge \dots \wedge \alpha^{2p} \, . 
 $$
 The noncommutative dimension ${\rm dim}(E)$ of  a module with such  charges can be evaluated via Elliott's formula 
\cite{Elliott}. One should 
apply to $\mu(E)$ an operator 
\begin{equation} \label{Elliott}
exp\left( -\frac{\partial}{\partial \alpha_{j}}\theta^{jk}\frac{\partial}{\partial \alpha_{k}} \right)
\end{equation}
and take the zero grading component. The condition for $\mu(E)$ to give topological numbers  of a 
projective module is ${\rm dim}(E)>0$. 
For example a system with $n$ D0 and $m$ D2p charges  defines a projective module if
$n + m{\rm Pf}\theta >0$. 
 In particular we may have negative $n$ or $m$ in some ranges  allowed by the value of ${\rm Pf}\theta$.    
In that case we cannot represent such a module as a direct some of D0 and D2p  modules and they have no solutions 
analogous to the solitons we have been considering so far. However  these modules can be represented  as a difference 
of modules in the 
sense of $K$-theory group. This suggests a physical interpretation of these  modules as the ones  describing a result of tachyon 
condensation of tachyons in   D2p-$\overline{\rm D0}$ or $\overline{\rm D2p}$-D0  systems. For example consider  
the situation when ${\rm Pf}\theta>0$ 
and a module has a noncommutative dimension $1-{\rm Pf}\theta$. 
It is a real (as opposed to a virtual one) module when ${\rm Pf}\theta<1$ or 
${\rm Pf}(B/(2\pi))>1$ that is when the induced D0 charge due to the magnetic field on the D2p-brane 
can compensate the charge $-1$ of the 
anti D0-brane. Tachyon condensation processes for   non-BPS branes and  brane-antibrane systems on noncommutative tori was 
studied in \cite{Bars}, \cite{Schn}, \cite{Mats}.  


\section{Some open problems} \label{discussion}
For D0-D2 and D0-D4 systems that  we discussed in detail 
we know the solitonic configuration describing a point-like 
D0 sitting on top of the higher-dimensional brane and we know that there exists on the same module an exact solution 
describing  a BPS configuration in which D0 is smeared  over the D2p brane. For D0-D2 system this is a constant curvature connection and 
for D0-D4 it is a noncommutative instanton. However we have not given an explicit construction of either of them. 
And the reason is that the direct sum representation of the module is not well suited for that purpose.  
Let us illustrate this point on the simplest example of D0-D2 system.   

 From general results on 
noncommutative tori \cite{RieffelProj} we know that any module over a two-dimensional torus (with irrational $\theta$)
 admits a constant curvature connection. One can choose 
a gauge in which an arbitrary constant curvature connection takes the form $\nabla_{j}^{cc} + c_{j}\cdot {\bf 1}$ where 
 $\nabla_{j}^{cc}$ is some fiducial constant curvature connection and $c_{j}$ are numbers. To find $\nabla_{j}^{cc}$ 
explicitly in the direct sum representation of our D0-D2 module we need to take a general connection in the form (\ref{c2})
and plug it into the equation $F_{12} = const\cdot {\bf 1}$.
In terms of components this is  a system of 4 coupled equations that so far resisted our attempts to solve it directly. 
   On the other hand there is another
   representation of the D0-D2 module in which it is very easy to construct such a connection. 
In this representation an element of $E$ is represented by a single function $f(x)\in  L_{2}({\mathbb R}^{1})$ and the 
torus generators act as 
$$
f(x)U_{1} = f(x-\theta -1)\, , \qquad f(x)U_{2} = e^{2\pi i x} f(x)
$$
and a constant curvature connection can be chosen as 
$$   
\nabla^{cc}_{1} = \frac{2\pi i}{\theta + 1}x \, , \qquad  \nabla^{cc}_{2}= \frac{\partial}{\partial x} \, .  
$$
We know that both representations of the D0-D2 module are isomorphic, the problem is to find an explicit isomorphism. 
One can write the following general ansatz for such an isomorphism $M:E_{0,1}\oplus E_{1,0}\to L_{2}({\mathbb R}^{1})$: 
\begin{eqnarray*}
&&M: (\psi(x) , U_{\bf n}) \mapsto S\psi(x) + \chi(x)U_{\bf n} \, , \\
&&M^{-1}: \phi(x) \mapsto (T\phi(x) , \langle \tilde \chi, \phi\rangle_{T_{\theta}})  
\end{eqnarray*}
where $S$, $T$ are some operators on $L_{2}({\mathbb R}^{1})$ and $\chi(x)$, $\tilde \chi(x)$ are some fixed functions. 
The condition that $M^{-1}$ in the second line is indeed the inverse to $M$ gives the following set of equations: 
\begin{eqnarray} \label{equations} 
&& S^{\dagger}\tilde \psi = 0\, , \qquad T\psi = 0 \, , \nonumber \\
&& TS=1 \, , \qquad ST = 1 - \langle \tilde \psi, \psi\rangle_{T_{1/\theta}}  
\end{eqnarray}
from which it is clear that $\langle \tilde \psi, \psi\rangle_{T_{1/\theta}}$ is a projector that cuts out of $L_{2}({\mathbb R}^{1})$ 
a submodule  isomorphic to $E_{0,1}$ and the complement submodule is a cyclic one spanned by the action of $T_{\theta}$ 
on a single vector $\psi(x)$. This construction is an analog of the partial isometry operators $S$ and 
$S^{\dagger}$  used to construct  solitons on the plane. 
So far we have been unable to find a suitable solution of these equations.

In \cite{unstable} 
a consistent truncation of massive modes was proposed based on the symmetry properties of both the solitonic solution and 
the solution describing the final state of the tachyon condensation. It was shown that the massive modes can  then be integrated out 
leading to an effective tachyon potential. For a similar analysis for the tachyon condensation on a torus one seems to need at 
least an explicit solution to (\ref{equations}). We leave this problem among others for a future investigation.

\begin{center}
{\bf Acknowledgments} \end{center}

I would like to thank Nikita Nekrasov  for useful discussions and the Erwin Schroedinger Institute for Mathematical Physics 
where  part of this work was done for a warm hospitality.

\end{document}